\documentclass[a4paper,12pt]{article}
\usepackage[dvips]{graphicx}
\begin{document}

\begin{center}
{SINGLE--PASS LASER POLARIZATION OF\\ 
ULTRARELATIVISTIC POSITRONS.}
\end{center}

\begin{center}
{ \large Alexander \ Potylitsyn }\\

{ \sl Tomsk Polytechnic University,}\\

{ \sl pr. Lenina 2A, Tomsk, 634050, Russia}\\
{\scriptsize \rm e-mail: pap@phtd.tpu.edu.ru}
\end{center}
\begin{abstract}
The new method for producing of the polarized
relativistic positrons is suggested. A beam
of unpolarized positrons accelerated up to a
few GeV can be polarized during a head-on
collision with an intense circularly polarized
lazer wave. After a multiple Compton backscattering
process the initial positrons may lose a substantial
part of its energy and, as consequence, may acquire
the significant longitudinal polarization. The simple
formulas for the final positron energy and
polarization degree depended on the laser flash
parameters have been obtained. The comparison of
efficiences for the suggested technique and known
ones is carried out. Some advantages of the new
method   were shown.
\end{abstract}

The experiments with polarized electron-positron beams in 
 future linear 
colliders  will furnish a means for studying 
a number of intriguing physical problems [1]. 
While the problem of generation and acceleration of 
longitudinally polarized  electron beams seems to be solved [2], 
the approach for production of polarized positron beams with 
the required parameters has not been finally defined yet.
 In [3-7] methods were proposed for  the generation of longitudinally
 polarized positrons during $\rm e^{+} e^{-}$-- pair production by circularly 
polarized photons with the energy $\rm \omega \sim 10^1$ MeV, which are, 
in their turn, generated by either passing electrons with the 
energy $\sim10^2$ GeV through a helical undulator [3], or through 
Compton backscattering of circularly polarized laser 
photons on a beam of electrons with the energy $\sim$5 GeV [4,5], 
or through bremsstrahlung of longitudinally polarized $\sim$50 MeV
electrons  [6,7]. To achieve the needed intensity of a positron 
source 
($\rm N_{e^+, pol}\sim10^{10}$ particles/bunch) it is suggested to use 
an undulator of the length $\rm L >$ 100 m [8], or to increase the 
laser power [9], or to use a high-current accelerator of 
polarized electrons [10]. 

The present paper considers an alternative way to approach this problem.

A beam of unpolarized positrons from a conventional source  being cooled 
in a damping ring and preliminary accelerated to an energy $\rm E_0$ can be 
 polarized during a head-on-collision with a high-intensity 
circularly polarized laser wave.

It is well known that during Compton backscattering of circularly 
polarized laser photons on unpolarized  positrons (electrons) with the 
energy E$_0\sim$100 GeV the scattered photon takes up to 90$\%$
 of the initial positron energy while the recoil positron 
aquires $\sim 100 \% $  longitudinal polarization [11,12]. 
At  $\rm E_0 \leq $10 GeV, however, the positron  
 loses too little of its energy during
single Compton backscattering (a few percent), and the longitudinal
 polarization of the recoil 
positron is, therefore, of the same order of magnitude. Current 
advances of laser physics make it possible to obtain parameters of
  laser flash such that the positron \underline {successively} interacts with 
$\rm N \gg$ 1 identical circularly polarized photons. It is apparent that 
in this case the positron can lose a substantial fraction of its energy 
(comparable with $\rm E_0$). To evaluate the resulting polarization of the 
recoil positron, let us consider multiple Compton backscattering 
in greater detail.

Let us carry out calculations in a positron rest frame (PRF)  
 and in a laboratory frame (LF).
 Following [13], let us write the Compton scattering cross section 
in PRF where spin correlations of three particles will be viewed-- 
initial photon, and initial and recoil positrons 
(upon summation over the scattered photon polarizations):
$${\rm \displaystyle{\frac{d\sigma}{d\Omega}= 2 r_0^2 
\Big(\frac{k}{k_0}\Big)^2
\Big\{\Phi_0 + \Phi_2(P_c, \vec\xi_0) + \Phi_2(P_c, \vec\xi) +
\Phi_2(\vec\xi_0, \vec\xi) + \Phi_3(P_c, \vec\xi_0, \vec\xi)\Big\}}}
\eqno(1)
$$
Here $\rm r_0$ is the electron classical radius; \
$\rm k_0$,\  $\rm k$ are the initial and 
scattered photon energy; \ 
$\rm P_c = \pm$ 1 is the circular polarization of 
the initial photon; \ 
and $\rm \vec{\xi_0}, \ \rm \vec{\xi} $ are the spin vectors of the initial 
and final positrons.
Functions $\rm{\Phi_0, \ \Phi_2, \ \Phi_3} $
were obtained in paper [13].

In (1) and further in the paper use is made of the system of units
$\rm{\hbar = m_e = c} = 1$, unless otherwise indicated. 

Since the scattered photons 
are not detected, the cross section (1) has to be integrated over 
the photon outgoing angles. Due to  azymuthal symmetry, 
it will depend on  the average longitudinal polarization components
 $\rm{\xi_{0l},  \ \xi_l}$ solely.
 On this basis we will keep only these components which remain 
 the same in LF.

For positrons with $\rm{ \gamma_0 \ \leq 10^4 \ (\gamma_0}$  is the 
Lorentz-factor of the initial 
positron), the laser photon energy in PRF ( $\rm{\omega_0 \sim }$1 eV in LF )
 will satisfy the relation
$$
\rm{ k_0 = 2 \gamma_0 \ \omega_0 \ll} 1 \eqno(2)
$$   
Using (2) let us write the expression for the scattered photon energy in PRF:
$$ \rm{ k = \frac{k_0}{1+k_0(1 - \cos \theta)} \approx k_0\big[1 - k_0 
(1 - \cos \theta)\big]}  
\eqno(3)
$$
Here $\rm\theta$  is the polar angle of the scattered photon  in PRF. 

Leaving the terms not higher than $\rm k^2_0$ , let us write in explicit 
form the $\rm \Phi_i$ functions derived in [13] for electrons :
$$ \rm{\Phi_0 = \displaystyle{\frac{1}{8}\Big[1 + \cos^2 \theta +
k^2_0 (1 -\cos \theta^2)\Big]}}\;, 
$$
$$
\rm{\Phi_2(P_c, \xi_{0l}) = \displaystyle{ - \frac{1}{8} P_c \xi_{0l}k_0 
\cos \theta}}\;,
\eqno(4)
$$
$$
\rm{\Phi_2 \big(P_c, \xi_{l}\big) = \displaystyle{-\frac{1}{8}} P_c \xi_{l}
\big(1-\cos\theta\big) 
 \Big[2k_0\cos\theta - k^2_0(\cos\theta - 
\cos^2\theta + \sin^2\theta)\Big]}\;,
$$
$$
\rm{\Phi_2(\xi_{0l}, \xi_l) =\displaystyle \frac{1}{8}\xi_{0l}\xi_l\big[1+ 
\cos^2\theta - k^2_0 \cos\theta\sin^2\theta\big] }\;,
$$
$$
\rm{\Phi_3\big(P_c, \xi_{0l}, \xi_l}\big)=0 \;.
$$

Upon routine integration we obtain:
$$\rm{
\sigma = \displaystyle{\frac{\pi r_0^2}{2}}
 \Big\{\displaystyle{\frac{8}{3}(1-2k_0)}+
\displaystyle{\frac{4}{3}}
P_c\xi_{0l}k_0(1-2k_0)+  
\xi_l\Big[\displaystyle{\frac{8}{3}}\xi_{0l}(1-2k_0)
 + \displaystyle{\frac{4}{3}}P_c k_0\Big]\Big\} }
\eqno(5)
$$
It is obvious that in averaging with respect to  
the  initial  particles spin and  taking  the
 summation with respect to two spin states 
of the recoil positron,  instead of (5) we get 
Klein-Nishina's cross section for $\rm k_0 \ll $ 1 [11]:
$$\rm{
\sigma = \frac{8}{3} \pi r^2_0 \big(1-2k_0\big)}
\eqno(6)
$$
From (5) follows that longitudinal polarization of a recoil 
positron (electron) is determined by both its initial polarization 
and the circular polarization of  a photon 
(later the longitudinal polarization indices $ l$ will be omitted):
$$\rm{
 \xi = \frac{\xi_0 \mp \displaystyle\frac{k_0}{2}P_c}{1 \mp
\displaystyle\frac{k_0}{2}P_c\xi_0} }
\eqno(7)
$$
The upper (lower) sing refers to a positron (electron).

If the initial positron is unpolarized ($\rm \xi_{0}$ = 0), 
 then upon a single interaction with a laser photon the recoil 
positron becames polarized :
$$ \rm{\vert
\xi_{(1)}\vert = \vert\frac{-k_0}{2}P_c\vert \ll} 1\;.
\eqno(8)
$$
In order to consider the next scattering act, let us calculate the 
average longitudinal momentum $\rm < k_{\parallel} >$ along the initial
 photon direction  and the average energy $\rm <k>$ of the scattered photon 
in PRF using the same approximation as before:
$$\rm{
 <k_\parallel> = \frac{ \int k \cos \theta 
\Big(\displaystyle\frac{k}{k_0}\Big)^2 \Phi_0 d\Omega }
 {\int \Big(\displaystyle\frac{k}{k_0}\Big)^2 \Phi_0 d\Omega} = 
 \displaystyle\frac{6}{5} k^2_0 }\;,
\eqno(9)
$$
$$\rm{
<k> = \frac{\int k \Big(\frac{k}{k_0}\Big)^2 \Phi_0 d\Omega}
{\int \displaystyle\Big(\frac{k}{k_0}\Big)^2 \Phi_0 d\Omega }
 = k_0 (1-k_0)}\;.
$$
Thus, upon the first event of interaction, the photon in LF aquires,
 on average, the energy
$$\rm{
< \omega_{sc} > = \gamma_0 (<k> - \beta_0 <k_{\parallel}>)
\approx \gamma_0<k> = \gamma_0 k_0} \;.
\eqno(10)
$$
In (10) $\rm{\displaystyle \beta_0 = 1- \frac{\gamma^{-2}_{0}}{2}}$ 
 is the velocity of PRF with respect to LF.

It is apparent that the recoil positron loses its energy (10) and hence
$$ \rm{
\gamma_{(1)} = \gamma_0 - <\omega_{sc}> = \gamma_0(1-k_0)=
\gamma_0(1-2\gamma_0\omega_0)}\; .
\eqno(11)
$$
In PRF before the second interaction the initial photon, in view of (11), 
will have a lower energy
$$\rm{
k_{(1)} = 2\gamma_{(1)} \omega_0 = 2\gamma_0 \omega_0
(1-2\gamma_0\omega_0) = k_0(1-k_0)} \;,
\eqno(12)
$$
and the recoil positron will have a  polarization: 
$$\rm{
\xi_{(2)} = \frac{\xi_{(1)} - \displaystyle\frac{k_{(1)}}{2}P_c}
{1 - \displaystyle\frac{k_{(1)}}{2} P_c\xi_{(1)}}} \;.
\eqno(13)
$$
Substituting its value from (8) for $\rm \xi_{(1)}$, we obtain:
$$\rm{
\xi_{(2)} = - P_c \frac{ \displaystyle\frac{k_0}{2} + 
\displaystyle\frac{k_{(1)}}{2}}
   { 1 - P_c \displaystyle\frac{k_0}{2}  \displaystyle\frac{k_{(1)}}{2}}  \;,
 \  \ \ \ \vert\xi_{(2)}\vert > \vert\xi_{(1)}\vert   }
\eqno(14)
$$
It follows from (14) that as a result of multiple Compton
 backscattering 
the longitudinal polarization of positrons builds up, while their energy 
decreases in LF (so-called laser cooling, see [14,\ 15]).

Let us write expressions relating the polarization and energy for 
two subsequent acts of scattering: 
$$\rm{
 \gamma_{(i+1)} = \gamma_{(i)}(1 - 2\omega_0\gamma_{(i)}})\;,
\eqno(15)
$$
$$\rm{
\xi_{(i+1)} = \frac{\xi_{(i)} - \gamma_{(i)} \omega_0 P_c}
{1 - \gamma_{(i)}\omega_0 P_C\xi_{(i)}} }\;.
\eqno(16)
$$
From these we can obtain the  equations for the finite differences:
$$\rm{
\Delta\gamma_{(i)} = \gamma_{(i+1)} - \gamma_{(i)} = 
2\omega_0\gamma^2_{(i)}}\;,
\eqno(17)
$$
$$\rm{
\Delta\xi_{(i)} = \xi_{(i+1)} - \xi_{(i)} \approx - \omega_0 P_c 
\gamma_{(i)}
(1-\xi^2_{(i)})}\;.
\eqno(18)
$$
When $\rm N \gg$1, instead of (17) and (18) we can arrive at differential 
equations, whose solution with proper initial conditions will yield
$$\rm{
\gamma_{\scriptscriptstyle(N)} = \frac{\gamma_0}{1 + 2\gamma_0\omega_0 N}}
 \eqno(19) 
$$
$$\rm{
\xi_{(N)} = \frac{\gamma_0\omega_0 N}{1+ \gamma_0\omega_0 N} }\;.
\eqno(20)
$$
When deriving the above relation, there was taken the left circular
polarization $\rm P_c =-1$   for the 
sake of simplicity.

Equations (19) and (20) describe the positron characteristics after 
$N$ collisions with circularly polarized laser photons. 
The number of collisions $\rm N$ is controlled by the luminosity of the process
$\rm L$:
$$\rm{ 
N = \frac{N_{scat}}{N_e^{+}} = N_0 L= N_0 \frac{\displaystyle
\frac{8}{3}\pi r^2_0 }
{2\pi(\sigma^2_{e^{+}} + \sigma^2_{ph})}}\;.
\eqno(21)
$$
In (21) $\rm N_0 = A / \omega_0$  is the number of photons per laser 
flash, $\rm A$ is the laser energy, and $\rm\sigma_{ph},  \ \rm\sigma_{e^+}$,
 are the laser focus and positron 
bunch radii. We can expect that after cooling in the damping ring 
$\rm{\sigma_{e^+} \ll \sigma_{ph}}$. In this case, substituting (21) 
into (19) and (20), 
we obtain the following simple formulas for positron's characteristics:
$$\rm{
 \gamma_{\scriptscriptstyle{ (N)}} = \frac{\gamma_0}{1 + 2\mu }}\;,
\eqno(22)   
$$
$$\rm{\xi_{(\scriptscriptstyle{N)}} = \frac{\mu}{1 + \mu}}\;,
\eqno(23)
$$
which depend on the  dimensionless parameter $\rm\mu$ solely
$$\rm{
\mu = \gamma_0 \omega_0 N = \displaystyle\frac{4}{3} \ 
\displaystyle\frac{A}{mc^2 }\gamma_0
\displaystyle\Big(\frac{r_0}{\sigma_{ph}}\Big)^2}\;.
\eqno(24)
$$
It follows from (24) that the $\rm\mu$ parameter depends linearly on the 
laser flash energy and the initial positron energy, but it is inversely 
proportional to the laser focus area and does not depend on the 
interaction time (duration flash). 
  Having written (22) as:
$$\rm{
\frac{\gamma_0}{\gamma_{\scriptscriptstyle{(N)}}} = 1 + 2 \ \mu}\;,
\eqno(25)
$$
we will compare the result with the estimate by V. Telnov [15] 
obtained in a classical approximation. Substituting into (24) the estimate 
used in [15] $\displaystyle 
\rm{\sigma^{2}_{ph}} = \frac{\rm \lambda_0 l_e}
{8\rm \pi}$ ( $\rm \lambda_0$ is the laser photon wavelength 
and $\rm l_e$ is the positron bunch length), we get :
$$\rm{
\frac{\gamma_0}{\gamma_{\scriptscriptstyle (N)}} = 1 + 
\frac{64}{3} \ \frac{A}{mc^2}
\ \gamma_0\frac{\pi r^2_0}{\lambda_0 l_e}}\;.
\eqno(26)
$$
  The resulting expression is closed to  a similar one in [15] 
but  the second term in (26) is by a factor of $\rm\pi$ smaller.
This dicrepancy is connected with rough calculation of the luminosity
(constant area of the laser focus) used in (21).

   By way of illustration let us consider an example (see [15]): 
$$\rm{
\gamma_0 = 10^4, \ A = 5 \ J, \ \lambda_0 = 500 \ nm, \ l_e = 0.2\ mm,\
\ \sigma^2_{ph} = \lambda_0 l_e / 8\pi}\;.
\eqno(27)
$$
 In this case $\rm \mu $= 1.6 and, therefore, 
 $\rm {\gamma_{\scriptscriptstyle(N)} \approx 
0.3 \gamma_0; \  \xi_l \approx 60\% }$.

 Thus, when a positron bunch interacts with a laser flash of 
the given parameters, \underline{all} the positrons acquire longitudinal 
polarization of about 60$\%$. The change in the polarization sign
 is obtained by inverting the  sign of the circular polarization 
of laser radiation.

  It should be noted that with a proper selection of the sign 
of circular polarization, the process of laser cooling would 
give rise to a longitudinal polarization increase of the 
electrons rather than to depolarization of electrons beam
(as in the case of unpolarized laser radiation considered in [15]).

  Note that, generally speaking, the laser parameters (27) 
correspond to the so-called "strong" field, when the 
contribution from non-linear Compton scattering [4] would be 
considerably high. 

Non-linear processes, i.e., simultaneous scattering of 
a few laser photons on the moving positron, are characterized 
by an increase in the effective positron mass  in PRF, which, 
in its turn, leads to a decrease in the Lorentz-factor 
and the energy transferred to the positron through scattering. 
It is to be expected that the $\rm\mu$ parameter (24) for a fixed 
value of the laser flash energy A will be sufficiently lower 
for a non-linear case as compared to the linear one, 
and hence a lower attainable polarization (23).

In order to reach a linear mode of the  Compton scattering process, 
one has  to 
stretch the laser flash (the length of its interaction with 
the positron bunch) (see, for instance,  [16]). 

In conclusion, let us estimate the energy $\rm {A_{+,pol}}$ necessary 
to obtain one polarized positron  with  the energy 
$\rm {E_{+}} > 10^1$  
MeV  and the longitudinal polarization $\rm \xi_l >$ 0,5 i.e., 
the parameters acceptable for consequent acceleration.

i) According to the estimates [8] an  electron with 
the energy $\rm{E_{-} \sim}$  200\ GeV  passing through a helical undulator of 
the length $\rm{L\sim 150\ m}$ can generate a number of circularly 
polarized photons needed to obtain one polarized positron 
 to be later accelerated (conversion efficiency 
$\rm{ \eta = \displaystyle{ N_{e^+, pol}}/{N_{e^{-}}} \approx} 1$). \ 
Hence, $\rm{A_{+,pol} {\sim} {E_{-}}/{\eta} = 200 \ GeV}$. 

ii) The author of the paper [9] considered a scheme for production
 of $\rm{\sim N_{e^+} = 10^9}$ polarized positrons when the 
laser radiation of the total energy $\rm{A_{ \Sigma} 
\sim 20\ J}$ is
scattered  on an 
electron bunch with $\rm E_{-}$ = 5 GeV and $\rm {N_{e^-}} = 10^{10}e^{-}$
/ bunch.
Thus
$$\rm{
A_{+,pol} \approx \frac{N_{e^-} E_{-} + A_{\scriptscriptstyle {\Sigma }}}
{N_{e^+}}
\approx 170 \ GeV} \;.  \nonumber
$$

iii) In [6] the author estimated the conversion  
efficiency for longitudinally 
polarized electrons  with the energy $\rm{E_{-}}$ = 50 MeV:
$$ \rm{
\eta \approx 10^{-3} }\;.$$
 Therefore $\rm{\displaystyle A_{+,pol} \sim {E_{-}}/\eta} =$ 50 GeV.

iv) For the method suggested in the present paper, 
evaluation of $\rm{A_{+,pol}}$ can be made for 
parameters of the  unpolarized positron source  
used in SLAC [17].

The conversion efficiency for the electron energy 
$\rm E_{-} $ = 33 GeV equals:
$$\rm{
\eta \approx} 1 \;. \nonumber
$$
Therefore, for a  bunch with $\rm{N_{e^+}} = 10^{10}$  
 and the positron energy $\rm{E_0}$ = 5 GeV interacting with the laser
flash  ($\rm A = 5\ J$) we have:
$$\rm{
\displaystyle A_{+,pol} = \frac{E_{-}}{\eta}+E_0 + \frac{A}{N_{e^{+}}}  = 
33\ GeV + 5 \ GeV + 3\ GeV \sim 40\ GeV}\;.
$$
Thus, the scheme proposed here seems to be most 
energy effective.

The author is grateful to V. Telnov and J. Clendenin 
for stimulating discussions.

\vspace*{10mm}
\centerline{\large\bf  References}
\begin{enumerate}
\item
P.M.\, Zerwas. Preprint  DESY 94-001, 1994.
\item
J.E.\, Clendenin, R.\, Alley, J.\, Frish, T.\, Kotseroglou,
G.\,Mulhollan, D.\,Schultz, H.\,Tang, J.\,Turner and
A.D.\, Yeremian. The SLAC Polarized Electron Source.
AIP Conf. Proceedings, N. 421, pp. 250-259, 1997.
\item
V.E.\, Balakin, A.A.\,Mikhailichenko. Preprint INP 79-85, 
Novosibirsk, 1979.
\item
Yung Su Tsai. Phys.\,Rev.\,D, v.48 (1993), pp.96-115.
\item
T.\, Okugi, Y.\, Kurihara, M.\, Chiba, A.\, Endo, R.\, Hamatsu,
T.\, Hirose, T.\, Kumita, T.\, Omori, Y.\, Takeuchi, M.\, Yoshioka.
Jap.J.Appl.Phys. v. 35(1996), pp. 3677-3680.
\item
A.P.\, Potylitsyn. Nucl. Instr. and Meth. v. A398 (1997), pp.395-398.
\item
E.G.\, Bessonov, A.A.\, Mikhailichenko. Proc. of V European 
Particle Accelerator Conference, 1996, pp. 1516-1518.
\item
K.\, Flottmann. Preprint DESY 93-161, 1993.
\item
T.\, Omori. KEK- Proceedings 99-12, 1999, pp. 161-179.
\item
T. \,Kotseroglou, V.\, Bharadwaj, J.E.\, Clendenin, S.\, Ecklund,
J.\, Frisch, P.\, Krejcik, A.V.\, Kulikov, J.\, Liu, T.\, Maruyama,
K.K.\, Millage, G.\, Mulhollan, W.R.\, Nelson, D.C.\, Schultz, 
J.C.\, Sheppard, J.\, Turner, K.\, Van Bibber, K.\, Flottmann,
Y.\, Namito.   Particle Accelerator Conference (PAC'99).
Proceedings (to  be published).
\item
H. \,Tolhoek. Rev. of Mod. Phys. v. 28 (1956), pp. 277-298.    
\item
G.I.\,Kotkin, S.I.\, Polityko, V.G.\, Serbo. Physics of Atomic Nuclei,
v. 59 (1996), pp. 2229-2234.
\item
F.W.\,Lipps, H.A.\,Tolhoek. Physika, v. 20 (1954), pp. 395-405.
\item
P.\,Sprangle, E.\,Esarey. Phys. Fluids. v.B4 (1992), pp. 2241-2248.
\item
V.\, Telnov. Phys. Rev. Lett. v.78 (1997), pp.4757-4760.
\item
I.V.\, Pogorelsky, I.\, Ben-Zvi, T.\, Hirose. BNL Report No.65907,
October, 1998.
\item
S.\, Ecklund. SLAC-R-502 (1997), pp. 63-98.      

\end{enumerate}

\end{document}